# High performance hard magnetic NdFeB thick films for integration into Micro-Electro-Mechanical-Systems

N.M. Dempsey, A. Walther, F. May and D. Givord

*Institut Néel, CNRS, 25 ave de Martyrs, 38042, Grenoble, France*

K. Khlopkov and O. Gutfleisch

*IFW Dresden, Institute of Metallic Materials, Helmholtzstr. 20, 01069 Dresden, Germany*

5μm thick NdFeB films have been sputtered onto 100 mm Si substrates using high rate sputtering (18 μm/h). Films were deposited at ≤ 500°C and then annealed at 750°C for 10 minutes. While films deposited at temperatures up to 450°C have equiaxed grains, the size of which decreases with increasing deposition temperature, the films deposited at 500°C have columnar grains. The out-of-plane remanent magnetization increases with deposition temperature, reaching a maximum value of 1.4 T, while the coercivity remains constant at about 1.6 T. The maximum energy product achieved (400 kJ/m$^3$) is comparable to that of high-quality NdFeB sintered magnets.

NdFeB thin films with excellent hard magnetic properties have been prepared by sputtering [1] and Pulsed Laser Deposition.[2] Out-of-plane texture was induced by depositing the films onto heated substrates in both 1-step (directly crystallized) [1,2] and 2-step (crystallized during a post-deposition anneal)[3,4,5] processes. Thick films (≥5 µm) of such high performance hard magnetic materials have potential applications in magnetic MEMS (Micro Electro Mechanical Systems).[6] Their integration into MEMS requires the use of preparation techniques which can produce high deposition rates over large areas. In addition, the use of substrates which are compatible with today's MEMS technology (i.e. Si) is necessary for the exploitation of such materials. So far, the only group which has reported the successful growth of textured NdFeB thick films by high rate sputtering (≈30 µm/h) used metallic substrates.[3] in this letter, we report on the preparation and analysis of 5µm thick NdFeB films deposited on 100 mm Si substrates using high rate sputtering. The film's magnetic properties compare with those of high-quality sintered NdFeB magnets.

{Ta (100 nm) / NdFeB(5 µm) / Ta (100 nm)} films were deposited onto (001) oriented Si substrates (diameter = 100 mm) at a rate of 18 µm/hour by triode sputtering of square targets of surface area 100 mm x 100 mm (note that the film thickness is reduced to 4 µm at the edge of the 100 mm wafer). The nominal composition of the NdFeB target was $Nd_{16.8}Fe_{74.7}B_{8.5}$. During deposition, the substrate was either "cold" (no power to substrate heater, though the substrate temperature gradually rises during deposition, eventually reaching a temperature of about 230°C) or fixed at a temperature in the range 300 – 500°C. Full wafers were annealed at 750°C for 10 minutes. Structural characterization was carried out using high resolution Scanning Electron Microscopy (LEO Gemini 1530, equipped with In-Lens and Quadrant Back Scattering (QBSD) detectors as well as Energy Dispersive X-ray Analyzer (EDX)) and x-ray diffraction (Co radiation). For cross-sectional imaging, the samples were simply fractured while for plane-view imaging they were lightly polished to remove the Ta capping layer. Room temperature magnetic properties were characterized with a Vibrating Sample Magnetometer.

In the as-deposited state, films prepared at $T_{sub} \leq 450°C$ appear to be amorphous (no x-ray diffraction peaks are resolved and the films are magnetically soft) while the films deposited at 500°C are crystallized (appearance of $Nd_2Fe_{14}B$ xrd peaks and an out-of-plane coercivity of 0.3T) (data not shown). Structural and magnetic properties of the annealed samples are compiled in figure 1. The samples' fractured cross-sections, observed with an in-lens detector, are shown in column 1. Films deposited at substrate temperatures up to 450°C have an equiaxed granular microstructure, in stark contrast with films deposited at 500°C, which have a columnar structure with individual grains traversing the entire film thickness of 5μm (note that in the as-deposited state, no grain structure is observed up to 450°C while a columnar structure exists in films deposited at 500°C, however it is less distinct than in the annealed state). Inter-granular fracture is predominant and many grain surfaces have a peculiar sponge like aspect due to the presence of small spherical holes ($\phi$ = 10-50 nm), the density of which decreases with increasing deposition temperature. Such holes were already identified in films prepared by a two-step process.[3,4] Kapitanov et al. suggested that they may be due to the segregation of trapped Ar during the post-deposition annealing process.[3]

The samples' polished plane-view sections, imaged with a Quadrant Back Scattering Detector, are shown in column 2. While electron backscattered diffraction analysis (not shown) identifies the grey grains in these plane-view images as $Nd_2Fe_{14}B$, no diffraction patterns could be obtained on the large dark zones observed. EDX analysis indicates that the films have an overall Fe:Nd atomic ratio of 6.1, which is slightly lower than the value of 7 for $Nd_2Fe_{14}B$, indicating that any significant quantity of additional phases should be Nd-rich and thus show a brighter contrast compared with $Nd_2Fe_{14}B$ grains in the QBSD images. TEM analysis (imaging, EELS and EDX) indicates that these dark zones are in fact holes.[7] The holes show a wide distribution in size, and except for the "cold" deposited sample, the size of the largest holes is comparable to that of the $Nd_2Fe_{14}B$ grains. Very few holes are observed in the films deposited at 500°C, and these are significantly smaller that the average diameter of the $Nd_2Fe_{14}B$ columnar grains. These large irregular shaped holes appear to be specific to our preparation procedure. Their size depends on the surface area of the annealed sample as well as the annealing conditions (heating rates). We attribute their formation to concurrent grain growth and volume change upon crystallization (a 10% increase in density has been reported upon crystallization of amorphous RE-TM films).[8] The direct crystallization of films deposited at 500°C explains the near absence of such holes in these films.

Refinement of the $Nd_2Fe_{14}B$ grain size with increasing deposition temperature up to 450°C is observed, with the average grain size decreasing from a value of the order of 800 nm for the sample deposited on a "cold" substrate to a value of the order of 200 nm for the sample deposited at 450°C. This indicates that the density of $Nd_2Fe_{14}B$ nucleation sites increases with substrate temperature. It is likely that these nucleation sites are responsible for the development of texture in the films, the exact nature of the nucleation sites remaining to be established (a study of films in their as-deposited state is underway). The change to a columnar structure for a deposition temperature of 500°C is attributed to the direct crystallization of $Nd_2Fe_{14}B$ during deposition under growth conditions favorable to columnar grain formation.

Though all films show out-of-plane crystallographic c-axis texture, a progressive increase in the degree of texture with increasing deposition temperature is evidenced by a significant sharpening of the (006) x-ray diffraction pole figures shown in the third column of figure 1. Correspondingly, an increase in out-of-plane magnetic anisotropy is clearly identified in the fourth column, in which in-plane and out-of-plane hysteresis curves are compared. Though other authors attributed the anisotropic magnetic nature of films produced by such a two-step procedure to the development of a columnar grain structure,[4] we clearly demonstrate that such a columnar structure is not necessary. The variation of remanence, coercivity and energy product with deposition temperature are plotted in figure 2. While the remanence increases with deposition temperature, due to the increase in out-of-plane texture, the coercivity varies little ($\mu_0 H_c \approx 1.6$ T). This is in contrast to bulk NdFeB samples, in which easy-axis coercivity tends to be reduced as texture is increased.[9] The maximum energy product (400 kJ/m$^3$) which was achieved for the films deposited at 500°C is comparable to that of high-quality sintered magnets.[10]

All films show a two-stage initial magnetization behavior in the out-of-plane measurements with a first stage of high susceptibility in low fields (<0.2T) followed by a second low susceptibility stage. The relative contribution of the former decreases with increasing deposition temperature up to 450°C, and then rises again for a deposition temperature of 500°C. We attribute this 2-stage behavior to the presence of both multi-domain (high susceptibility) and single-domain (low susceptibility) Nd$_2$Fe$_{14}$B grains.[11] The influence of deposition temperature on the relative contribution of each population to the sample's magnetization is directly related to its influence on grain size. The switch from single to multi-domain state should occur at the critical single domain particle size, which for Nd$_2$Fe$_{14}$B is estimated to be 300 nm.[12] This agrees with the experimental observation that the biggest change in the relative contribution of low and high susceptibility behavior occurs between 300°C and 400°C, when the average grain size is refined from above to below 300 nm. The rise in multi-domain behavior in the film deposited at 500°C may be explained by an increase in grain volume owing to the shift from equiaxed to columnar grains which traverse the entire film. An alternative explanation for the low susceptibility stage would be domain wall pinning. However, the irregular shaped holes present in our films are much too large to act as pinning sites (to be effective, defects should be of a size comparable to the domain wall width, which for $Nd_2Fe_{14}B$ is estimated to be 5 nm),[13] while the density of nanometer sized spherical pores which could act as efficient pinning sites decreases as the low susceptibility contribution increases.

In addition to the anomalous magnetic volume effect (Invar effect), Kapitanov et al. demonstrated that the thermal expansion curves of NdFeB films produced by such a two step procedure display an expansion anomaly during crystallization of the $Nd_2Fe_{14}B$ phase.[3] Owing to this, they separated their films from their metallic substrates, prior to annealing them in small pieces. In this study, where the post-deposition annealing was carried out on the entire 100 mm wafer, it was found that the central region (diameter ≈ 30 mm) of all films has a tendency to peel off during annealing (note that all results presented above were for sections of film which remained tightly adhered to the Si substrate). We attribute this effect to a combination of three factors: the above mentioned expansion anomaly during crystallization, differential thermal expansion of the magnetic layer and the substrate and to a degradation of the Si - Ta interface caused by diffusion during the high temperature annealing step.[7] This peel off is not necessarily problematic since the out of plane magnetization of the films is optimally exploited in structured objects of which the lateral dimensions are of the order of the film thickness. No peel off occurs in Ta (100 nm)/ NdFeB (5μm) / Ta (100 nm) films deposited through a mask with mm sized holes or onto pre-patterned 100 mm wafers which have topographic relief (trench motifs with individual trenches of depth 5 μm with trench/wall widths on the scale of 5-100 μm).[14] We have recently produced mechanically intact 50 μm thick films on 100 mm Si wafers by depositing through the same mask and using a 1-step procedure in which the films are directly deposited in the crystalline state.

We have studied the structural and magnetic properties of thick NdFeB films deposited on Si substrates. Though the best textured films have a columnar structure, such a structure is not necessary for magnetic anisotropy. The development of texture in films produced using a two step procedure is due to the formation of oriented nucleation sites during the initial deposition step. The optimum films produced have magnetic properties ($\mu_0 M_r = 1.4$ T, $\mu_0 H_c = 1.6$ T and $(BH)_{max} \approx 400$ kJ/m$^3$) which are comparable to those of the best sintered bulk NdFeB magnets. The combination of high sputtering rates, large sputtering surfaces, micro-technology compatible substrates and excellent magnetic properties are very promising for the integration of hard magnetic materials into MEMS.

This work was carried out in the framework of the ANR "Nanaomag2" project. The authors would like to thank N. Kornilov for his invaluable contribution to the triode sputtering activity at Institut Néel. K. Khlopkov is grateful for financial support from the DFG (SFB 463).

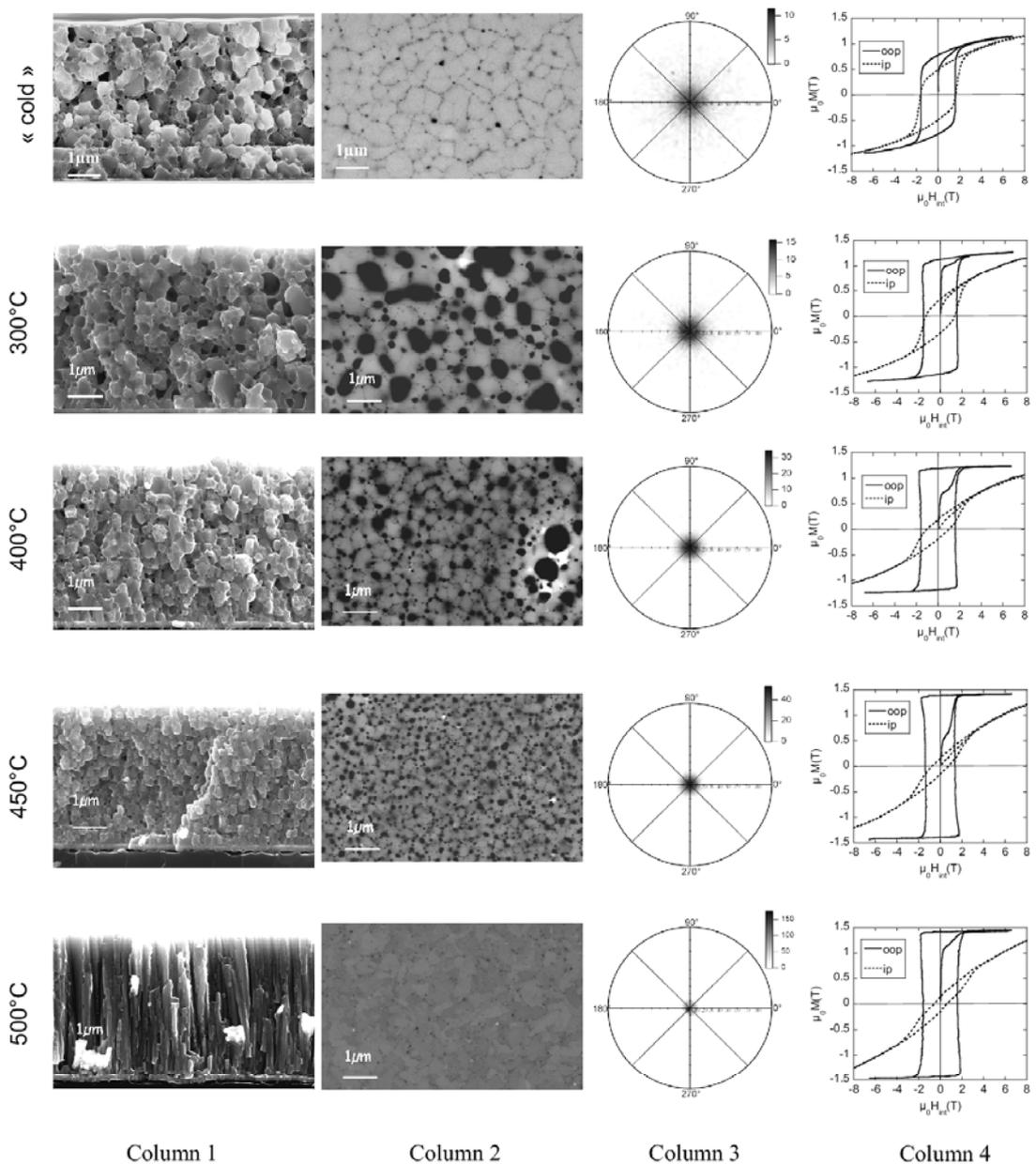

FIG. 1. Comparison of the evolution in structural and magnetic properties of NdFeB films as a function of deposition temperature (note all films were annealed at 750°C for 10 minutes).
Column 1 : Fractured cross sections imaged with an in-lens detector
Column 2 : Polished plane views imaged in QBSD mode
Column 3 : XRD pole figure of (006) diffraction peak of $Nd_2Fe_{14}B$
Column 4 : In-plane (ip) and out-of-plane (oop) hystersis loops and the initial oop magnetization curves (oop loops were corrected with a demagnetization factor N = 1 ).

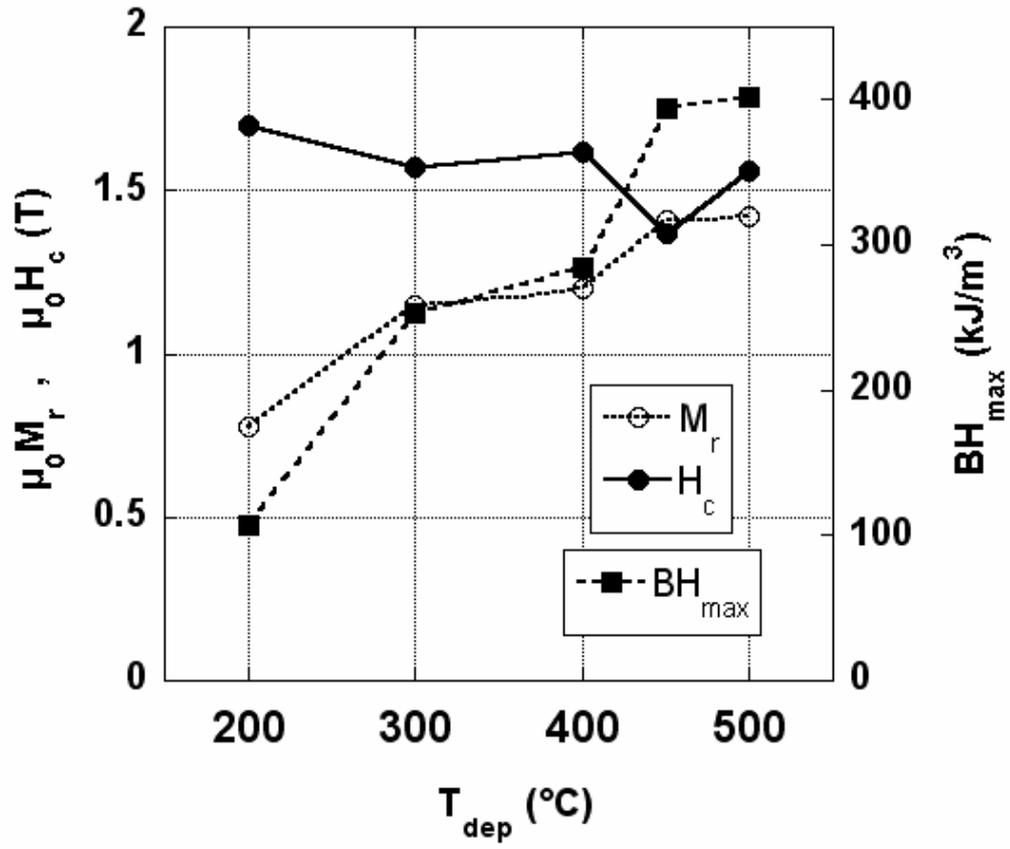

FIG. 2. Remanence $\mu_0 M_r$, coercivity $\mu_0 H_c$ and estimated maximum energy product $(BH)_{max}$ as a function of deposition temperature (note all films were annealed at 750°C for 10 minutes).